\documentstyle[aps,prl,graphicx,floats]{revtex}
\begin{document}
\draft
\twocolumn[\hsize\textwidth\columnwidth\hsize\csname @twocolumnfalse\endcsname

\title{ Thermodynamics of the superfluid dilute Bose gas with disorder.}

\author{A.V. Lopatin  and  V.M. Vinokur}
\address{
Material Science Division, Argonne National Laboratory,
Argonne, Illinois 60439
}
\date{\today}
\maketitle

\begin{abstract}
We generalize the Beliaev-Popov diagrammatic technique for the
problem of interacting dilute Bose gas with weak disorder. Averaging
over disorder is implemented by the replica method. Low energy
asymptotic form of the Green function confirms that the low energy
excitations of the superfluid dirty Boson system are sound waves with
velocity renormalized by the disorder and additional dissipation
due to the impurity scattering. We find the thermodynamic potential
and the superfluid density at any temperature below the superfluid
transition temperature and derive the phase diagram in temperature 
vs. disorder plane.

\end{abstract}

\vskip2pc]

\bigskip

Superfluidity in random environments enjoys a long standing yet
intense attention. The effect of disorder on the behavior of
systems possessing long-range correlations is central to
contemporary condensed matter physics, and superfluid Bose gas
offers an exemplarily unique and accessible tool for both
experimental and theoretical researches. One of the fascinating
properties of such systems is their ability to maintain
superfluidity (i.e. long range correlations) even in the strongly
disordered environment. He$^4$ , for example, remains superfluid
when absorbed in porous media \cite {Experiments}. The problem of
influence of disorder on superfluidity (and on its close analog
-superconductivity) has been under extensive theoretical attack
(see seminal works \cite{theory1,Fisher}) and remarkable progress
in qualitative understanding of disordered Bose systems was
achieved. Recent papers \cite{Huang,Pitaevskii} discussed a
continuum model of the dilute interacting Bose gas in a random
potential. The advantages of this model are that (i) it is
microscopically related to the original problem and (ii) it is
very well understood in the clean limit. The proposed model
describes, in particular, the quasiparticle dissipation and
depletion of superfluidity at zero temperature and marked an
important step towards quantitative description of disordered Bose
systems.

In this Letter, building on the model of
Refs.\cite{Huang,Pitaevskii}, we develop a systematic diagrammatic
perturbation theory for the dilute Bose gas with weak disorder at
finite temperatures below the superfluid transition temperature
$T_s$. We obtain disorder corrections to the thermodynamic
potential which completely determine thermodynamic properties of
the superfluid system. We derive for the first time the
disorder-induced shift of $T_s$ resulting from disorder scattering
of quasiparticles with energy $\epsilon\sim T$. We find that the
superfluid density decreases monotonically with the temperature.
This completely agrees with the experimental data, while being in
some contradiction with the theoretical result of Ref.\cite{Huang}
where a non-monotonic temperature dependence of superfluid density
dependence was claimed. In the limit $T\to 0 $ our theory
reproduces all the results of Refs.\cite{Huang,Pitaevskii}.

{\it The model.} The starting point of our model is the Lagrangian density:
\begin{equation}
{\cal L}=-\varphi ^{*}(-\nabla _r^2/2m-\mu +u(r)+\partial _\tau )\varphi
-g\;\varphi ^{*}\varphi ^{*}\varphi \,\varphi ,
\end{equation}
where $\varphi =\varphi (r,\tau )$ is the field representing Bose particles,
$r$ is the real space coordinate, $\tau $ is the Matsubara time and $u(r)$
is the disorder potential. As usual we consider a soft interaction potential
$g(r)$ and use the Born approximation $g=4\pi \lambda /m$ to relate the
interaction constant $g=\int g(r)d^dr$ to the scattering length $\lambda $.
\cite{Yang} Taking Gaussian ${\delta}$-correlated disorder $\langle
u(r_1)u(r_2)\rangle =v\,\delta (r_1-r_2)$ we derive the effective replicated
action in a form:
\begin{equation}
S=-\sum_p\varphi _\alpha ^{*}(p)(k^2/2m-\mu -i\omega )\varphi _\alpha
(p)+V_i+V_d,
\end{equation}
with $p=(k,\omega )$ and the interaction parts $V_i$ and $V_d$
\begin{eqnarray}  \nonumber
V_i=-\frac g{2\beta V}\sum_{k,\omega ,\alpha }\varphi _\alpha
^{*}(k_{1,}\omega _1)\varphi _\alpha ^{*}(k_2,\omega _2)\varphi
_\alpha (k_3,\omega _3)\varphi _\alpha (k_4,\omega _4)\\
\nonumber
 V_d=\frac \kappa {2V}\sum_{k,\omega ,\alpha ,\beta
}\varphi _\alpha ^{*}(k_{1,}\omega _1)\varphi _\alpha (k_3,\omega
_1)\varphi _\beta ^{*}(k_2,\omega _2)\varphi _\beta (k_4,\omega
_2)
\end{eqnarray}
where $\alpha,\beta =1,...,n$ are the replica indices, $\kappa
=v^2$ and the conservation of total momentum $k_1+k_2=k_3+k_4$ 
(in $V_i$ and $V_d$) and of total 'energy' $\omega _1+\omega
_2=\omega _3+\omega _4$ (in $V_i$ ) is assumed. The corresponding
vertices are presented in Fig.1a. Below $T_s$  we separate the
condensate contribution by shifting the fields $\varphi _\alpha
\rightarrow a\sqrt{ \beta V}\delta _{k,0}\delta _{\omega
,0}+\varphi _\alpha ^{\prime }$ and define the Green functions of
the fields $\varphi _\alpha ^{\prime }$ by
\begin{equation}
G_{\alpha \beta }(p)=\left\langle \varphi _\alpha ^{\prime }(p)\varphi
_\beta ^{\prime *}(p)\right\rangle, F_{\alpha \beta }(p)=\left\langle
\varphi _\alpha ^{\prime }(p)\varphi _\beta ^{\prime }(-p)\right\rangle .
\end{equation}
Defining also the functions $\bar G_{\alpha \beta }(p)=\langle \varphi
_\alpha ^{\prime *}(-p)\varphi _\beta ^{\prime }(-p)\rangle ,$ $\bar F%
_{\alpha \beta }=\langle \varphi _\alpha ^{\prime *}(-p)\varphi _\beta
^{\prime *}(p)\rangle $ we introduce the matrix Green function and the
corresponding matrix self energy
\begin{equation}
{\cal G}_{\alpha ,\beta }=\left[
\begin{array}{cc}
G_{\alpha \beta } & F_{\alpha \beta } \\
\bar F_{\alpha \beta } & \bar G_{\alpha \beta }
\end{array}
\right] ,\;\;\;\;\Sigma _{\alpha \beta }=\left[
\begin{array}{cc}
A_{\alpha \beta } & B_{\alpha \beta } \\
\bar B_{\alpha \beta } & \bar A_{\alpha \beta }
\end{array}
\right],
\end{equation}
that are related by the Dyson equation
\begin{equation}
\label{Dyson}{\cal G}^{-1}=(p^2/2m-\mu -i\omega \tau _3)\delta _{\alpha
\beta }+\Sigma ,
\end{equation}
where $\tau _3$ is the Pauli matrix. The condensate density is uniformly
distributed in the replica space, therefore Green function ${\cal G}$ and
self energy $\Sigma $ can be presented as
\begin{equation}
\label{GSigma}{\cal G}_{\alpha \beta }={\cal G}_1\,\delta _{\alpha \beta }+
{\cal G}_2\,R_{\alpha \beta },\;\;\;\Sigma _{\alpha \beta }=\Sigma
_1\,\delta _{\alpha \beta }+\Sigma _2\,R_{\alpha \beta }
\end{equation}
where $R_{\alpha \beta }$ is the matrix with all elements equal to $1$. From
the $n\rightarrow 0$ limit of Eq.(\ref{Dyson}) we obtain that the replica
diagonal part of the Green function is determined by the replica diagonal
part of the self energy
\begin{equation}
\label{g1g2}{\cal G}_1^{-1}=(p^2/2m-\mu -i\omega \tau _3)+\Sigma _1,\;\;
{\cal G}_2=-{\cal G}_1\Sigma _2{\cal G}_1,
\end{equation}
and the poles of the Green function ${\cal G}$ are determined by the poles
of the Green function $G_1.$ Due to the Goldstone theorem, below the
condensation temperature the function $G_1$ has a pole at $p=0$, and from (%
\ref{g1g2}) we find
\begin{equation}
\label{PGT}A_1(0)-B_1(0)=\mu.
\end{equation}

\begin{figure}[ht]
\includegraphics[width=3.2in]{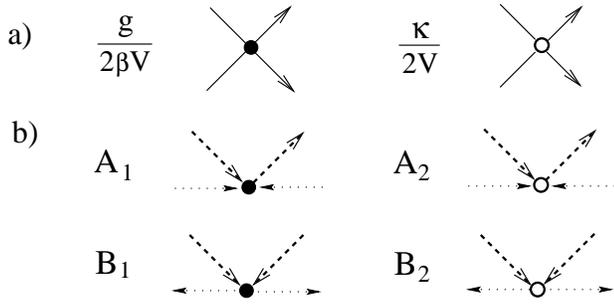}
\vspace{0.3cm} \caption{Interaction and disorder vertices (a) and
first order contributions to self energies (b). Dashed lines in
(b) represent the condensate. }
\end{figure}
Our diagrammatic technique parallels that by Beliaev, the difference being
that we include the disorder vertex $V_d$ along with the interaction vertex $%
V_i.$ The corresponding diagrams are shown in Fig 1b.
\begin{equation}
\label{FirstOrder}A_1=2ga^2,\quad B_1=ga^2,\quad A_2=B_2=-\kappa \beta a^2.
\end{equation}
From Eq.(\ref{PGT}) we obtain $ga^2=\mu $ and using Eq.(\ref{g1g2}) for the
Green functions we have
\begin{eqnarray}
\label{G1F1}G_1(p)=\frac{k^2/2m+\mu +i\omega }{\varepsilon ^2(k)+\omega ^2}%
,\quad F_1(p)=\frac{-\mu }{\varepsilon ^2(k)+\omega ^2}, \\
\label{G2F2}G_2(p)=F_2(p)=\frac{\kappa \,\beta a^2}{(k^2/2m+2\mu )^2}\delta
_{\omega ,0},
\end{eqnarray}
where $\varepsilon (k)=\sqrt{(k^2/2m)^2+\mu k^2/m}.$ We see that
the spectrum of quasiparticles is not affected by disorder in the
leading order. The Bose gas density is given by
\begin{equation}
\label{density1}
n=a^2+n_1+n_2
\end{equation}
with
\begin{equation}
\label{density2}
n_1={\frac{T}{V}} \sum_{p\,}G_1(p), \quad n_2={\frac{T}{V}}
\sum_{k,\omega =0}G_2(k).
\end{equation}
{\it Zero temperature.} At zero temperature in three dimensions
using Eqs.(\ref{G1F1},
\ref{G2F2} ) we obtain
\begin{equation}
\label{n1n2_0}
n_1=\frac 8{3\sqrt{\pi }}(\lambda n)^{3/2},\quad n_2=\frac
\kappa {4\pi }\frac{a^2m^{3/2}}{\sqrt{\mu }}.
\end{equation}
The contribution $n_1$ represents the quasiparticle density due to quantum
fluctuations, in the leading order it coincides with the well known answer
for the pure case. The contribution $n_2$ represents the density of
nonuniform part of the condensate. For the theory to be applicable both $n_1$
and $n_2$ should be much less than the total density
\begin{equation}
\label{Condition}\lambda ^{\prime }=\lambda n^{\frac 13}\ll 1,\quad \kappa
^{\prime }=\kappa \,m^2/(8\pi ^{\frac 32}\sqrt{\lambda n})\ll 1.\quad
\end{equation}
The first relation is the usual low-density condition, the second one
insures that the uniform part of the condensate is not strongly affected
by disorder. To relate the condensate density with the chemical potential we
need to improve the leading order result $ga^2=\mu $ considering next order
corrections to (\ref{PGT}). The second order corrections $%
A_1^{(2)},B_1^{(2)} $ contain the contributions $A_1^{(2,i)},B_1^{(2,i)}$
from the quasiparticle interactions and the disorder contributions $%
A_1^{(2,d)},B_1^{(2,d)}$ linear in $\kappa .$ Corrections $
A_1^{(2,i)},B_1^{(2,i)}$ presented in Fig 2a exactly coincide with ones
studied in \cite{Beliaev} for the pure Bose gas. The disorder corrections $%
A_1^{(2,d)},B_1^{(2,d)}$ presented in Fig 2b contain all the diagrams that
are (i) linear in disorder coupling $\kappa $ and (ii) have a similar
structure with ones shown on Fig 2a. Here we present the answers for their
linear combinations $\Sigma _{\pm }^{(2,d)}=A_1^{(2,d)}\pm B_1^{(2,d)}$
\begin{eqnarray}
\label{RSigma-}
Re\,\Sigma _{-}^{(2,d)}(q,\omega )&=&gn_2-{\kappa\over V} \sum_k\frac{k^4\,G_{-}^{\prime }(k+q,\omega )\,}{(k^2+4\mu m)^2}, \\
\nonumber
Re\,\Sigma _{+}^{(2,d)}(q,\omega )&=&3gn_2-{\kappa\over V}
 \sum_k\frac{(k^2-8\mu
m)^2\,G_{+}^{\prime }(k+q,\omega )}{(k^2+4\mu m)^2},
\end{eqnarray}
where $G_{\pm }^{\prime }(p)=Re(G_1(p)\pm F_1(p)).$ The self energy $B$ is
real while $A$ is complex with
\begin{equation}
\label{ImA}Im\,A^{(2,d)}(q,\omega )=\sum_k\frac{\kappa \,\omega \,k^2/V}{
\varepsilon ^2(k+q)+\omega ^2}\frac{8\mu m-k^2}{(k^2+4\mu m)^2}.
\end{equation}
\begin{figure}[ht]
\includegraphics[width=3.2in]{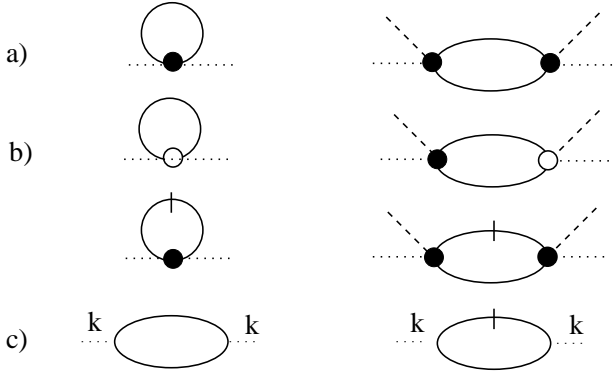}
\vspace{0.3cm}
\caption{Second order contributions to the self energies $A_1, B_1$ due
to interaction (a), and due to disorder (b). The solid lines represent
the Green functions $G_1, F_1$ the crossed solid lines represent the Green
functions $G_2,F_2.$ The diagrams in part (c) represent two contributions
to the normal density $n_n^{(1)}$ and $n_n^{(2)}. $}
\end{figure}
Using Eq.(\ref{RSigma-}) along with the well known result
\cite{Beliaev} for $\Sigma _{-}^{(2,i)}$ from Eq.(\ref{PGT}) we
obtain
$$
\mu =ga^2(1+\frac{40}3\sqrt{\frac{n\lambda ^3}\pi })+gn_2- {\frac{\kappa}{V}}
\sum_k \frac{2\,m\, \,k^2}{(k^2+4\mu m)^2}.
$$
Combining this equation with Eq.(\ref{density1}) we obtain the
relation between the density and chemical potential
\begin{equation}
\label{mu2}\mu =gn(1+\frac{32}3\sqrt{\frac{n\lambda ^3}\pi })-{\frac{\kappa}{%
V}} \sum_k \frac{2m\,k^2}{(k^2+4\mu m)^2}.
\end{equation}
Using the thermodynamic relation $\mu=dE/d N $ along with the leading order
relation $g n=\mu$ we obtain the energy
\begin{equation}
\label{Energy}
{{E}\over{V}}
=\frac{gn^2}2\left(1+\frac{128}{15}\sqrt{\frac{\lambda ^3n}\pi
} \right)- {\frac{\kappa}{V}} \sum_k\frac{2\,n\,m}{k^2+4\mu m},
\end{equation}
which agrees with \cite{Pitaevskii}. The integrals over $k$ in Eqs.(\ref{mu2}
,\ref{Energy}) are ultraviolet divergent, it is the consequence of the white
noise assumption for the disorder correlation function. This divergence is
not relevant for the low energy physics since it could be absorbed in the
renormalization of energy and chemical potential: $E \to E+n\kappa
\sum_k\,2m/ k^2,$ $\mu \to \mu+\kappa \sum_k\,2m/V k^2.$

The superfluid density $n_s$ can be found from the normal density $n_n$
which is determined by the transverse current-current correlator $n_s=n-n_n.$
In the leading order $n_n$ is given by the diagrams shown in Fig. 2c
\begin{equation}
\label{normalden}n_n=n_n^{(1)}+n_n^{(2)},
\end{equation}
where $n_n^{(1)}$ is the normal density of the clean system $n_n^{(1)}=\frac
T{3mV}\sum_p k^2(G_1^2(p)-F_1^2(p) ) $ which after summation over Matsubara
frequencies may be written as
\begin{equation}
\label{nnormal1}n_n^{(1)}={\frac{{1}}{{12 mTV}}}\sum_k{\frac{{k^2 }}{{%
\sinh^2[\epsilon(k)/2T]}}}
\end{equation}
and $n_n^{(2)}$ is the disorder correction
\begin{equation}
\label{nnormal2}n_n^{(2)}=\frac T{3mV}\sum_p2k^2[G_1(p)-F_1(p)]G_2(p)={\frac
4 3}n_2.
\end{equation}
Thus at zero temperature $n_n^{(1)}=0$ and superfluid density becomes $%
n_s=n-4n_2/3$ in agreement with Refs.\cite{Huang,Pitaevskii}.

The second order corrections to the self energies
(\ref{RSigma-}-\ref{ImA}) can be calculated explicitly for small
$q,\omega$ leading to the following low energy retarded Green
function
\begin{equation}
G_1^R(k,\omega)=\frac m{n_s}\, 
\frac{c^2\,a^2 }{c^2\,k^2-\omega ^2- 2ick\Gamma (k)%
},\;\;\; \omega>0
\end{equation}
where the sound velocity $c$ is related to the sound velocity of the clean
system $c_0$ by $c^2=c_0^2(1+5n_2/3n),$ and $\Gamma (q)=\kappa
\,\,q^4/24 m^2c^3 \pi $ is the dissipation of quasiparticles due to disorder
scattering. The dissipation due to quasiparticle (phonon) scattering is
known to be of a higher power of $q$: $\Gamma ^{(ph)}(q)\sim q^5$. The
results for sound velocity $c$ and quasiparticle dissipation $\Gamma(q)$ are
in agreement with \cite{Pitaevskii}.

{\it Bose condensation temperature.} Now we turn to finite
temperatures. At temperatures above the condensation temperature
$T_c $ the self energy is given by the first diagrams of Figs.
2a,b.
\begin{equation}
\label{HighTempA}A_1=2gn-{\kappa\over
V} \sum_kG(k,\omega ),
\end{equation}
and $A_2,B_{1,}B_2=0$. Taking the sum over $k$ in (\ref{HighTempA}) we
obtain the Green function
\begin{equation}
\label{HighTempG}G^{-1}(p)=\frac{k^2}{2m}-\widetilde{\mu }-i\omega +\kappa
\frac{(2m)^{\frac 32}}{4\pi }\sqrt{|\widetilde{\mu }|-i\omega },
\end{equation}
where the $\widetilde{\mu }=\mu -2gn+\kappa \sum_k2m/V k^2$. The density at $%
T=T_c$ can be easily obtained from Eq.(\ref{HighTempG}) taking into account
that at this temperature $\widetilde{\mu }$ becomes zero
\begin{equation}
\label{density_T_lam}n=\zeta _{\frac 32}\,(mT_c/\,2\pi )^{\frac 32}+\kappa
\,T_c m^3/4\pi ^2.
\end{equation}
Solving this equation for $T_c$ we find the shift of the condensation
temperature due to disorder:
\begin{equation}
\label{T_lambda}T_c =T_c ^{(0)}\left( 1-\kappa T_c ^{(0)} m^3/ 6\pi
^2n\right) ,
\end{equation}
where $T_c ^{(0)}=2\pi (n/\zeta _{\frac 32})^{\frac 23}/m$ is the
Bose condensation temperature of the ideal gas with $\zeta _{\frac
32}\approx 2.612$. At finite $g$ only microscopic amount of
particles may condense into a local potential well, this effect
leads to the smearing of the condensation transition making $T_c$ to be
the crossover temperature between the normal phase and a phase
where bosons are locally condensed. The true phase transition
takes place when the chemical potential reaches the mobility
edge\cite{theory1}, it may be also obtained from the condition
$n_s=0.$ (see below)

{\it Thermodynamics at} $T\sim T_s.$ The self energies at $T\sim T_s$  
are still given by the diagrams presented in Figs. 1 and 2,
but the first diagram of Fig.1a should be included already in the
first order approximation since the density of quasiparticle
excitations $n_1$ at $T\sim T_s$ is of the order of total density.
This diagram results only in the shift of the chemical potential
$\mu \to \mu -2gn_1=\bar \mu ,$ and the Green functions are still
given by the Eqs.(\ref{G1F1},\ref{G2F2}) but with $\mu \to \bar
\mu .$ The density in the leading approximation is still given by
Eqs.(\ref{density1},\ref{density2}), but now $n_1$ is not a small
correction and should be therefore calculated with a higher
accuracy. The main contribution to $n_1$ comes from the energies
$\epsilon \sim T\gg \bar \mu .$ At these energies the Green
function is given by Eq.(\ref {HighTempG}) and, thus, the disorder
correction to $n_1$ is the same as in Eq.(\ref{density_T_lam}),
i.e. $\kappa Tm^3/4\pi ^2.$ Taking into account this correction
and using Eq.(\ref{density2}) we obtain
\begin{equation}
\label{n_1HighT}n_1=
n T^{\prime \frac 32}-{\frac{{\bar \mu ^{\frac 12}m^{\frac
32}T}}{{2\pi }}}+{\frac{{\kappa \,m^3T}}{{4\pi ^2}}},\,\,n_2={\frac{{\kappa
\,a^2\,m^{\frac 32}}}{{4\pi \sqrt{\bar \mu }}}},
\end{equation}
where $T^{\prime }=T/T_c^{(0)}.$ To relate the chemical potential with the
condensate density we need to consider the next order corrections to the
leading order result $ga^2=\bar \mu $ following from Eq.(\ref{PGT}). Using $%
\Sigma _{-}^{(2,i)}$ found in \cite{Popov} along with Eq.(\ref{RSigma-}) we
obtain
$$
{\frac{{\bar \mu }}{{g}}}=n_2+a^2-\sum_k{\frac{{(\kappa /g)2mk^2}}{{(k^2+4%
\bar \mu m)^2}}}-{\frac{{3\bar \mu ^{\frac 12}m^{\frac 32}T}}{{2\pi }}}+{%
\frac{{\kappa m^3T}}{{2\pi ^2}}}.
$$
and combining this equation with Eq.(\ref{n_1HighT}) we get an
equation relating density and chemical potential
\begin{equation}
n=\mu /g-nT^{\prime \frac 32}+\bar \mu ^{\frac 12}m^{\frac 32}T/\pi +\Delta
n^{(d)},
\end{equation}
where $\Delta n^{(d)}$ is the disorder contribution
\begin{equation}
\Delta n^{(d)}={\frac{{\kappa }}{{gV}}}\sum_k{\frac{{2mk^2}}{{(k^2+4\bar \mu
m)^2}}}-{\frac{{\kappa \,m^3\,T}}{{4\pi ^2}}}.
\end{equation}
Using the relation $N=-d\Omega /d\mu $ we eventually obtain the disorder
correction to the  thermodynamic potential:
\begin{equation}
{{\delta \Omega ^{(d)}}\over{V}}
=-{\kappa\over{ Vg}}
\sum_k{\frac{{2\bar\mu m}}{{k^2+4\bar \mu m}}}
+{\frac{{\kappa m^3T\mu }}{{4\pi ^2}}}.
\end{equation}

{\it Superfluid density at $T\sim T_s.$} The disorder contribution
to the normal density $n_n^{(2)}$ at $T\sim T_s$ is related to
$n_2$ through Eq.(\ref{nnormal2}) with $n_2$ defined by
Eq.(\ref{n_1HighT}) that takes into account the chemical potential
shift $\mu \to \bar \mu.$ Deriving the contribution $n_n^{(1)}$
one needs to consider that according to Eq.(\ref{HighTempG}) the
spectrum of quasiparticles is affected by the disorder at energies
$ \epsilon \sim T_s,$ that results in
\begin{equation}
n_n^{(1)}=n_n^{(cl)}+\kappa m^3T/4\pi ^2.
\end{equation}
where $n_n^{(cl)}$ is the normal density of the clean system defined by Eq.(
\ref{nnormal1}) with the spectrum $\epsilon (k)=(k^2/2m)^2+\bar \mu k^2/m.$
Introducing dimensionless condensate density $a^{\prime }=\sqrt{1-T^{\prime
\frac 32}}$ we write the superfluid density as
\begin{equation} \label{sup_density}
n_s/n=n_s^{(cl)}/n-4\kappa ^{\prime }a^{\prime }/3-4\kappa ^{\prime }\sqrt{
\pi \lambda ^{\prime }}\,T^{\prime }/\zeta _{\frac 32}^{2/3},
\end{equation}
where $n_s^{(cl)}$ is the superfluid density of the clean system $
n_s^{(cl)}=n-n_n^{(cl)}.$ The dependence of $n_s$ on temperature
for different amounts of disorder is presented in Fig 3. Taking
$n_s=0$ in Eq.(\ref{sup_density}) we obtain the dependence of the
superfluid transition temperature $T_s$ on disorder as shown on
the phase diagram presented in Fig.3. The dotted line in this
diagram represents the condensation crossover temperature
determined by Eq.(\ref{T_lambda}).

\begin{figure}[ht]
\includegraphics[width=3.2in]{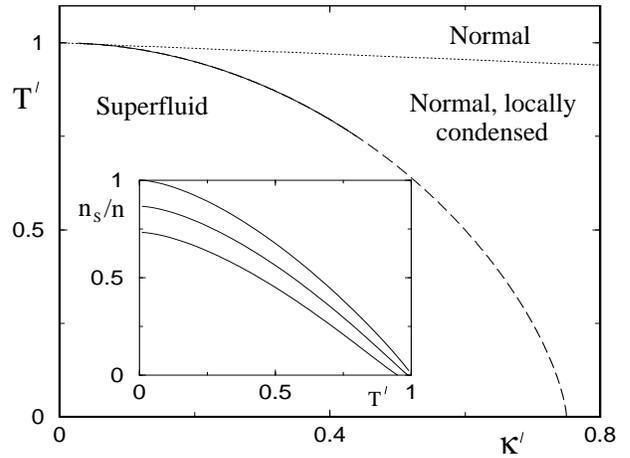}
\vspace{0.3cm} \caption{The temperature-disorder phase diagram resulting
from Eqs.(\ref{T_lambda},\ref{sup_density}) for
$\lambda^\prime=0.03^2.$
The dashed part of the boundary of the superfluid phase
corresponding to the region $\kappa\sim 1$ should be understood as an 
extrapolation. The insert shows
the superfluid density $n_s/n$ dependence on temperature
$T^\prime=T/T_c^{(0)}$ for different amounts of disorder:
$\kappa^\prime=0, 0.1, 0.2$ (top to bottom). }
\end{figure}

In conclusion, we have developed a regular diagrammatic approach that
enables a quantitative description of thermodynamics of superfluid dilute
Bose gas in random environment at finite temperatures. We have found
disorder corrections to condensation temperature, thermodynamic potential,
and the superfluid density. Our results agree favorably with the
experimental findings.

We would like to thank Lev Ioffe for useful discussions. This work
was supported by the U.S. Department of Energy, Office of Science
under contract No. W-31-109-ENG-38.

\end{document}